\documentclass[aps,prd,twocolumn,preprintnumbers,nofootinbib]{revtex4}
\usepackage{graphicx}

\begin{document}
\preprint{hep-th/0412197~~~UCB-PTH-04/36}
\title{Cosmology and the S-matrix}
\author{Raphael Bousso}\email{bousso@lbl.gov}
\affiliation{%
Center for Theoretical Physics, Department of Physics,
University of California, Berkeley, CA 94720, U.S.A.;\\
Lawrence Berkeley National Laboratory, 
Berkeley, CA 94720, U.S.A.}
\begin{abstract}
  We study conditions for the existence of asymptotic observables in
  cosmology.  With the exception of de~Sitter space, the thermal
  properties of accelerating universes permit arbitrarily long
  observations, and guarantee the production of accessible states of
  arbitrarily large entropy.  This suggests that some asymptotic
  observables may exist, despite the presence of an event horizon.
  Comparison with decelerating universes shows surprising
  similarities: Neither type suffers from the limitations encountered
  in de~Sitter space, such as thermalization and boundedness of
  entropy.  However, we argue that no realistic cosmology permits the
  global observations associated with an S-matrix.
\end{abstract}
\pacs{04.60.-m, 04.70.Dy, 11.55.-m, 95.30.-k}
\maketitle

\section{Introduction}

One problem with quantum gravity is that we don't know what the theory
should compute.  In particle physics, the most precise observable is
the S-matrix.  But this quantity seems ill-suited to cosmology, where
the observer is not outside the system, initial states cannot be set
up, and experiments cannot be arbitrarily repeated to gain
statistically significant results.

This ignorance is not especially unusual or embarrassing.  It is
rarely clear at the outset what a theory should compute.  For example,
the insight that gravity is a theory of a symmetric,
diffeomorphism-invariant tensor field in itself already constituted a
significant part of the development of general relativity.  But once a
theory is in its final form, the observables should be apparent.

If string theory is the correct quantum theory of gravity, then
whatever it computes presumably are the observables.  But string
theory---perhaps because it is {\em not\/} in its final form---has so
far sidestepped the problem of cosmological observables.  It defines
quantum gravity for certain classes of geometries characterized by
asymptotic conditions, such as asymptotically flat or Anti-de~Sitter
spacetimes.  In these geometries an S-matrix happens to make sense,
and string theory computes its matrix elements.  (In the case of AdS,
it computes boundary correlators, which are a close analogue of the
S-matrix.)

However, we have yet to learn how to apply string theory to cosmology
or to an observer inside a black hole, with the same level of rigor as
in Anti-de Sitter space.  Hence, it would be premature to conclude
that the S-matrix will remain the only well-defined object.  It is too
early to know what, if anything, string theory has to say about
cosmological observables.

Fortunately, classical and quantum properties of cosmological
solutions impose significant constraints on possible observables, and
may even hint at some of the principles on which a theory computing
them must be based.  De~Sitter space is a case in point.
Semi-classical analysis has provided overwhelming evidence that {\em
  no\/} exact observables exist in eternal de~Sitter space---at least,
none that correspond to experiments that can be performed by an
observer inside the universe.  This is related to the presence of a
cosmological event horizon in de~Sitter space, which limits the
accessible information and emits pernicious thermal radiation.

In this paper we use similar semi-classical reasoning to characterize
constraints on exact observables in other cosmological solutions.
Does the universe contain regions where fluctuations, including those
of the gravitational field, become arbitrarily weak?  Accurate
measurements take a long time\footnote{For this reason, we shall use
  the terms ``asymptotic observable'' and ``exact observable''
  interchangeably.}, and they require devices with a large number of
states.  Does the universe last long enough, i.e., does it contain
geodesics of infinite proper time in the future?  Does the causally
accessible region have enough quantum states?  According to entropy
bounds~\cite{Tho93,Sus95,CEB1}, this translates into a minimum
size for the region.  By asking whether such requirements are met, one
can investigate whether exact quantum mechanical observables exist in
a given cosmology, without knowledge of the full theory.

By an observable we mean a quantity or limit of quantities that can
actually be measured by an observer inside the universe, without
violating laws of physics such as causality or entropy bounds.  For
example, it may turn out that an S-matrix for de~Sitter space can be
formally computed as a useful ``meta-observable''~\cite{Wit01}, from
which predictions for true, operationally defined observables can be
extracted by further processing.  The restrictions derived below apply
only to the latter, operationally meaningful quantities.

Our conclusions for different classes of universes vary in their
details, but they do strike two common chords.  First: Aside from
de~Sitter space and the obvious case of crunching universes, our
necessary conditions for exact observables are satisfied in all the
other cases considered.  Surprisingly, this includes universes with a
cosmological event horizon.  Second: Observables that invoke a global
out-state (such as the S-matrix) do not seem to describe any
experiment in cosmology.  We find that the information content of a
generic out-state is causally inaccessible even in a universe with a
null infinity and no horizon.

We present a number of intermediate results that are of interest in
their own right: an analysis of the thermodynamics and the fluctuation
spectrum of quintessence universes; an argument demonstrating that an
open universe resides inside the Farhi-Guth solution; and entropic
reasoning suggesting that the global state of a non-compact universe
is not accessible to experiment, independently of event horizons.

This paper does not tackle the actual definition of any asymptotic
cosmological observables (see Ref.~\cite{BanFis01a,FreSus04} for
recent approaches).  Even that challenge, in turn, will only be an
intermediate goal.  In our view, asymptotic observables are at best a
crutch.  The description of a real experiment involving gravity
requires well-defined (but necessarily imprecise?) local observables.
This is a famously difficult problem in the presence of gravity.  It
is further complicated, but perhaps also helpfully constrained, by the
counter-intuitive holographic restriction on bulk degrees of
freedom~\cite{Bek81,Tho93,Sus95,FisSus98,CEB1,CEB2,RMP,Bou03}.  This
task will have to be confronted eventually.

\paragraph*{Relation to other work}

For a review of the difficulties with physics in de~Sitter space, see,
e.g., Ref.~\cite{Bou02b}.  A broad discussion of the problem of
observables in cosmologies with a non-positive cosmological constant
was given by Banks and Fischler~\cite{BanFis01a}, who noted that in a
non-compact universe, an S-matrix description must restrict to states
with a finite number of extra particles, and that those states are
very special.  While this restriction is necessary, it is not
sufficient: as shown below, the unobserved region can have infinite
entropy even if no particles are added, because of the internal states
of the matter already present.

Our analysis of the thermodynamics of Q-space builds on
Refs.~\cite{HelKal01,FisKas01}, who derived its global structure and
pointed out that its event horizon obstructs the definition of an
S-matrix.  We do not question this conclusion; indeed, we find that
the difficulties with an S-matrix are quite general in cosmology.  We
do argue, however, that other asymptotic observables may exist in
Q-space.  This possibility was first raised by Witten~\cite{Wit01},
who noted that observers will not be thermalized in Q-space.  The
existence of accessible high entropy states was not demonstrated
there.

The problem of defining cosmological observables is closely related to
the challenge of describing physics from the point of view of an
observer falling into a black hole.  In both cases, some type of local
observables will eventually be required, but in both cases, one can
hope to make progress by asking how some of the information in the
gravity-dominated region may be encoded in asymptotic
data~\cite{KraOog03,FidHub04}.

The recent discovery that the universe is accelerating has turned the
cosmological constant problem into the (worse) problem of small
positive vacuum energy.  Its possible resolution by a discretuum of
meta-stable vacua in string theory~\cite{BouPol00}, populated by
cosmological dynamics, makes it all the more urgent to understand
string theory observables in cosmology.  Explicit constructions of
de~Sitter vacua have been proposed (e.g., Ref.~\cite{KKLT}), and
sophisticated counting arguments (e.g.,
Refs.~\cite{AshDou03,GirKac04}) broadly confirm the original estimates
of the vast number of such vacua.  The present discussion does not
address specifically the development of a theoretical
framework~\cite{FreSus04,Ban04,BouFre05} describing this
``landscape''~\cite{Sus03}.  But the question of observables is a part
of this challenge, so our results may have some implications in this
context.

\paragraph*{Outline}

The paper is structured as follows.  In the first sections we mainly
consider spatially flat FRW universes with fixed equation of state
$w=p/\rho$.  They are especially simple and suffice for deriving our
main results.  Moreover, their late time behavior is a good
approximation to other classes of cosmologies, including some we
discuss at the end of the paper.

Sec.~\ref{sec-frw}, aside from a review of the flat FRW solutions and
their causal structure, contains our main observation about
decelerating universes ($w>-1/3$): All observers, at all times, lack
information about infinitely large regions of the universe, even
though there is no event horizon.  If such regions contain any
non-redundant information, then the global out-state
computed by an S-matrix cannot be measured.

Next, we turn to eternally accelerating universes, de~Sitter space
($w=-1$) and ``Q-space'' ($-1<w<-1/3$) \cite{RatPee88}\footnote{For a
  subset of this range, quintessence has been proposed as a model of
  dark energy~\cite{WanCal99}.  Here we study these solutions simply
  as instructive examples to understand conditions for asymptotic
  observables.}, which have a cosmological event
horizon~\cite{HelKal01,FisKas01}.  We show in Sec.~\ref{sec-temp} that
Q-space exhibits thermodynamic properties similar to those of the
de~Sitter horizon.  The horizon radius in Q-space grows linearly with
time, and consequently the temperature slowly decreases.  We find that
this behavior is consistent with the first law of thermodynamics: the
temperature and entropy respond appropriately to the flux of
quintessence stress-energy across the horizon.

Sec.~\ref{sec-fluc} contains our main results for accelerating
universes.  They support the existence of asymptotic observables in
Q-space.  We study specific aspects of the thermal spectrum emitted by
the horizon.  The time-dependence of the temperature leads to
significant differences between de~Sitter space and Q-space.  In the
semi-classical theory, an infinite number of Hawking quanta are
produced (and re-absorbed) by the horizon.  In de~Sitter space, the
total energy thus emitted diverges, whereas in Q-space the energy per
quantum decreases rapidly enough to render the total energy finite.
Hence, observers in Q-space will not be thermalized.

We ask whether observers will be destroyed by rare massive
fluctuations, such as black holes.  We consider objects of fixed
energy and compute the rate at which they are emitted, according to
standard statistical mechanics.  If the energy is much larger than the
temperature, the rate will be miniscule.  However, in de~Sitter space
the rate is constant, so all fluctuations that are not completely
forbidden will occur.  This guarantees that any observer who survives
the thermal radiation long enough will eventually be swallowed by a
large black hole emitted by the horizon.  In Q-space, the rate of such
violent processes decreases exponentially with time.  The integrated
probability is therefore finite and can be exceedingly small.  It
follows that experiments in Q-space can last for an arbitrarily long
time.

But the classical supply of matter in Q-space is bounded, seemingly
ruling out exact measurements.  Yet, we show in
Sec.~\ref{sec-qentropy} that arbitrarily complex matter configurations
are quantum mechanically {\em produced\/} by the Q-space horizon: The
rate for a fluctuation of a given fixed entropy---no matter how
large---is constant and non-vanishing at late times.  This contrasts
pleasantly with de~Sitter space, where the entropy is strictly bounded
by the inverse of the (fixed) cosmological constant.

In Sec.~\ref{sec-discussion} we draw conclusions on the nature of
observables in the universes we have studied.  In particular, we argue
that no direct analogue of an S-matrix can be defined in any flat FRW
universe unless the set of allowed states is severely restricted.

In Sec.~\ref{sec-other} we extend the discussion to open and closed
FRW solutions.  We also study composite universes that feature an
asymptotically flat region on the far side of a black hole.  We show
that the Farhi-Guth solution can be regarded as an example of this
setup in which the black hole resides inside an open universe produced
by the decay of meta-stable de~Sitter space.  Because an open universe
has infinite entropy, one would not expect generic microstates to be
represented on the far side of the black hole, or on the asymptotic
boundary.

\section{Spatially flat universes}
\label{sec-frw}

In this section we review various classical properties of flat FRW
universes---in particular, the results of
Ref.~\cite{HelKal01,FisKas01} on the causal structure of accelerating
cosmologies.  We ask how much matter and information is causally
accessible to an observer in the classical evolution.  We find that
this amount is finite in accelerating universes and unbounded in
decelerating universes.  However, even in the latter case, no more
than an infinitely small fraction of the matter is ever observable.

\subsection{Metric and causal structure}

\begin{figure}
\includegraphics[width=8.5cm]{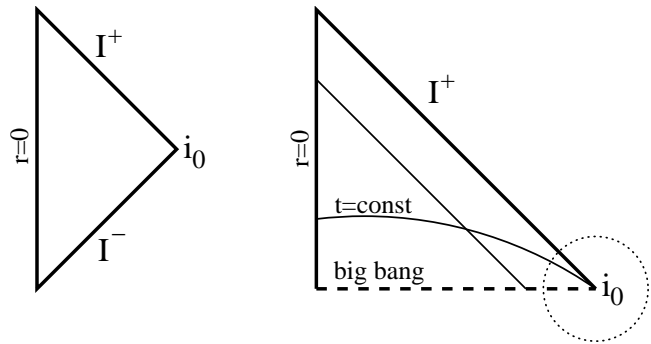}
\caption{\label{fig-penfrw} Conformal diagrams of Minkowski space (left) and a
  decelerating flat FRW universe (right).  In the FRW case, any
  infinitesimal neighborhood of spatial infinity (circle) contains an
  infinite amount of matter and potentially an infinite amount of
  information, whereas the observer's causal past is a finite region.}
\end{figure}

The metric of a spatially flat FRW universe is given by
\begin{equation}
ds^2 = -dt^2+a(t)^2 (dr^2+r^2 d\Omega^2) ~.
\label{eq-flatfrw}
\end{equation}
A quick way to obtain its causal structure is to transform to
conformal time, defined by $d\eta=dt/a(t)$.  This shows that the
metric is conformal to Minkowsi space: $ds^2 = a(\eta)^2
d\tilde{s}^2$, where
\begin{equation}
d\tilde{s}^2 = -d\eta^2+dr^2+r^2 d\Omega^2 ~.
\end{equation}
Hence, the conformal diagram is a subset of the Minkowski space
Penrose diagram (Fig.~\ref{fig-penfrw}), selected by the range of
$\eta$.  The existence of horizons is determined as follows.

If and only if $\eta$ is bounded from above ($\eta\to\eta_{\rm
  max}<\infty$ as $t\to\infty$), then an observer at $r=0$ is
surrounded by a future event horizon.\footnote{By homogeneity, all
  comoving observers are equivalent, so we consider an observer at
  $r=0$.  Any non-comoving observer whose spatial position remains
  finite at late times has the same horizon as a comoving observer
  located at the same asymptotic spatial position.}  The horizon is
located at $r=\eta_{\rm max}-\eta$.  Light-rays originating beyond
this hypersurface never reach the observer.  Similarly, if $\eta$ is
bounded from below, then there exists a past horizon.\footnote{This
  assumes that the FRW solution in question is past inextendible.
  Hence this analysis does not apply to the flat slicing of de Sitter
  space.}  Events beyond this horizon cannot be influenced by the
observer.

The dynamical evolution of the scale factor and the matter density is
determined by the equations
\begin{eqnarray} 
\frac{\dot{a}^2}{a^2} & = & \frac{8\pi\rho}{3}\ ,\\
\frac{\ddot{a}}{a} & = & -\frac{4\pi}{3}(\rho+3p)\ . \label{eq-rho}
\end{eqnarray}
We will assume that the energy density, $\rho$, and pressure, $p$,
obey the equation of state
\begin{equation}
p=w\rho
\end{equation}
with constant $w$.  It will be more convenient to work with the
parameter
\begin{equation}
\epsilon = \frac{3}{2}(w+1)\ .
\end{equation}
Thus one obtains a family of solutions
parameterized by $\epsilon$,
\begin{eqnarray} 
a(t) & = & t^{1/\epsilon} ~, \label{eq-a} \\
\rho(t) & = & \frac{3}{8\pi \epsilon ^2 t^2} ~.
\end{eqnarray} 
except for $\epsilon=0$, which corresponds to a cosmological constant
$\Lambda$.  In that case a solution is given by
$a(t)=\exp[(\Lambda/3)^{1/2}t]$, $\rho=\Lambda/8\pi$.

We assume the dominant energy condition, which restricts $\epsilon$ to
the range $0\leq \epsilon\leq 3$.  From Eq.~(\ref{eq-rho}) we can see
directly that for $\epsilon>1$, the expansion of the universe
decelerates: $\ddot{a}<0$.  This includes the familiar cases of matter
domination ($\epsilon=3/2$) and radiation domination ($\epsilon=2$).
For $\epsilon<1$, on the other hand, the scale factor grows
increasingly rapidly: $\ddot{a}>0$.  The degenerate case $\epsilon=1$
will not be considered here.

As discussed more generally above, we transform to conformal time,
\begin{equation}
\eta=\frac{\epsilon}{\epsilon-1}\ t^\frac{\epsilon-1}{\epsilon} ~,
\end{equation}
to reveal the causal structure.  For decelerating universes
($\epsilon>1$), this expression shows that conformal time is bounded
below but unbounded above.  There is no future event horizon.  The
conformal diagram is given by the upper half ($\eta>0$) of the Penrose
diagram of Minkowski space (Fig.~\ref{fig-penfrw}).

\begin{figure}
   \includegraphics[width=8.5cm]{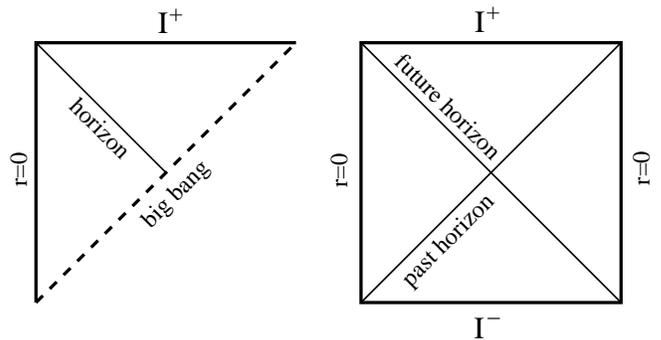}
\caption{\label{fig-penfrwb}
  Conformal diagrams of flat Q-space~\cite{HelKal01,FisKas01} (left)
  and de~Sitter space (right).  Past and future cosmological event
  horizons are shown.  The area of the de~Sitter horizon is constant,
  whereas the area of the Q-space horizon grows without bound at late
  times~\cite{HelKal01}.  The Q-space initial singularity is not
  really null, since the curvature already becomes Planckian on a
  nearby spacelike slice (see Fig.~3).}
\end{figure}

For accelerating universes with $0<\epsilon<1$, the situation is
reversed.  Conformal time ranges from $-\infty$ to $0$, and so is
bounded above but not below.  Hence, the conformal diagram is the
lower half of the Minkowski wedge (Fig.~\ref{fig-penfrwb}).  There is
a future event horizon at $r+\eta=0$, whose area, $A_{\rm E}$, grows
quadratically with time. The proper horizon area-radius, $R_{\rm
  E}=(A_{\rm E}/4\pi)^{1/2}$, is given by
\begin{equation}
R_{\rm E} = -\frac{\epsilon}{\epsilon-1}\ t~.
\label{eq-re}
\end{equation} 

In the case of de Sitter space, $\epsilon=0$, the metric (\ref{eq-flatfrw})
is geodesically incomplete and extendible.  The maximal extension has
closed spatial slices and is given by
\begin{equation}
ds^2 = \frac{3/\Lambda}{\sin^2\eta} (-d\eta^2 + d\chi^2 + \sin^2\chi
d\Omega^2)~.
\label{eq-global}
\end{equation}
Hence, the conformal diagram is a square (Fig.~\ref{fig-penfrwb}), and
de Sitter space has both past and future event horizons of constant
radius $\sqrt{3/\Lambda}$.

\subsection{Classical observable matter content}
\label{sec-class}

The maximum spacetime region probed by an experiment is called the causal
diamond~\cite{Bou00b}.  It is generally defined as the causal past of
the future endpoint of the observer's worldline, intersected with the
causal future of the past endpoint.  (Note that the latter is crucial:
events lying in the observer's past but outside the bottom cone cannot
be probed directly and may not send any signals in the observer's
direction.  If a signal is sent, then what information can be gleaned
about the event is precisely what passes through the bottom cone.)

How much matter enters an observer's causal diamond?  We restrict for
now to the classical evolution of the cosmological fluid, and postpone
the inclusion of the thermal properties of the horizon until
Sec.~\ref{sec-fluc}.

\paragraph*{de~Sitter space}

In eternal de Sitter space ($\epsilon=0$), the causal diamond is the
region limited by both the past and future event horizons.  The
maximum amount of matter that can enter is the largest black hole
allowed in asymptotically de Sitter space, the Nariai black hole.  Its
entropy is one third that of the empty de Sitter horizon.  But to
arrange for matter to enter, one must either include thermal effects,
or set up appropriate initial conditions in the infinite past.

\begin{figure}
  \includegraphics[width=8.5cm]{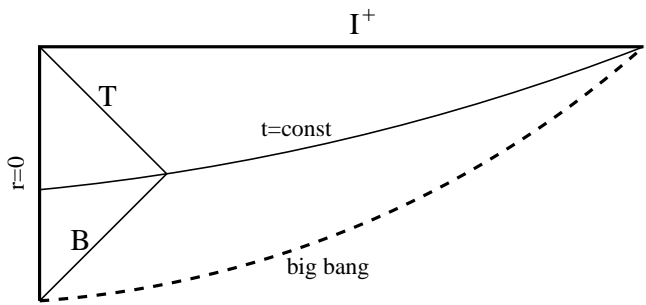}
\caption{\label{fig-penfrwc}
  This conformal diagram can be interpreted in three ways.  It
  represents pure Q-space, with a spacelike singularity reflecting a
  Planck scale cutoff of the classical metric (see Fig.~2).  It also
  corresponds to a big bang universe initially dominated by matter or
  radiation, which asymptotes to Q-space or de~Sitter space at late
  times.---The causal diamond of the observer at $r=0$ is shown.  The
  bottom cone (B) has finite maximal area, indicating that only a
  finite amount of entropy enters the observable region by classical
  evolution.  In asymptotically Q-space, however, the top cone (T)
  allows arbitrarily large entropy.  Indeed, an unbounded number of
  states can be accessed by quantum fluctuations of the horizon
  (Sec.~4.4).}
\end{figure}

It is more interesting to consider a universe such as ours, which
contains an era of matter- or radiation-domination before the
cosmological constant takes over.  The Penrose diagram for this type
of solution is shown in Fig.~\ref{fig-penfrwc}.  In that case, the
bottom cone of the causal diamond is the future light-cone of a point
at the big bang (usually called the particle horizon).  Its structure
depends on the details of the matter content.  But as long as the
universe is asymptotically de~Sitter in the future, the amount of
information inside the causal patch is bounded by the entropy at late
times, which is that of empty de Sitter space.

To summarize, an observer in asymptotically de~Sitter space can access
at most an entropy of order the inverse cosmological
constant~\cite{Ban00}.  This conclusion is independent of whether
thermal effects are included, and may extend to a larger class of
universes with positive cosmological
constant~\cite{Bou00a}.\footnote{If the requirement of a future
  asymptotic region dominated by the vacuum energy is dropped,
  examples with greater entropy are known in more than four spacetime
  dimensions~\cite{BouDew02}.}

\paragraph*{Q-space}

In an accelerating universe with $\epsilon>0$, the largest possible
causal diamond is the intersection of the past of the point $t=t_{\rm
  late}$, $r=0$ with the future light-cone of the point $t=1$, $r=0$,
in the limit $t_{\rm late}\to\infty$ (Fig.~\ref{fig-penfrwc}).  (We
follow Ref.~\cite{KalLin99} in excising the high curvature region
prior to the Planck time.  This replaces the null singularity with a
more standard, spacelike big bang.)  The lower cone is again the
particle horizon.  The upper cone, in the limit taken, is the future
event horizon (and so is a cone only conformally).

The amount of information entering the causal diamond from the past,
$S_{\rm in}$, is bounded by the maximal area of the lower
cone~\cite{FisSus98,CEB1}.  One thus finds that
\begin{equation}
S_{\rm in} \leq \pi \left(\frac{\epsilon}{1-\epsilon}\
2^{\frac{\epsilon}{1-\epsilon}}\right)^2~.
\end{equation}
Unless $\epsilon$ is very close to $1$, this is at most of order
unity, indicating that virtually no information enters the observer's
causal diamond.  Note that this result applies strictly to an
accelerating $\epsilon>0$ fluid with no other matter present.

The conclusion changes somewhat if other types of matter dominate at
early times.  If quintessence were the source of the vacuum energy in
our universe, for example, our particle horizon would intersect our
future event horizon about now (Fig.~\ref{fig-penfrwc}).  Its maximal
area would be quite large: about $10^{123}$ in Planck units.  Still,
like in de Sitter space, and unlike the decelerating universes, only a
finite amount of matter and information ever enter the causal diamond
by conventional evolution~\cite{GudBjo01,Loe01,KalKle04}.  To show
that Q-space exhibits unbounded complexity, one needs to include
thermal fluctuations (Sec.~\ref{sec-fluc}).

\paragraph*{Decelerating FRW}

In a decelerating universe, the bottom cone extends all the way to
future infinity and has infinite maximal area.  There is no bound on
the entropy that can enter.  Indeed any comoving particle {\em will\/}
enter it sooner or later.  Thus, in decelerating universes any
observer has access to arbitrarily large amounts of matter and
entropy.

However, there is an important order-of-limits issue.  Let us ask how
much of the universe is seen by an observer at some finite time $t$.
One finds that the sphere at the edge of the causal diamond has area
\begin{equation}
A_{\rm edge}=\pi\left(\frac{\epsilon t}{\epsilon-1}\right)^2~.
\label{eq-finite}
\end{equation}
This area is a bound on the amount of entropy that has entered the
region observed by the time $t$.  Note that this does not diverge at
finite $t$.  But finite $t$ is all an observer can ever attain.

Hence, the number of accessible degrees of freedom is, at all times,
an infinitely small fraction of the total number of degrees of freedom
in the universe.  This is shown in Fig.~\ref{fig-penfrw}; only the
past light-cone is shown (rather than the stronger restriction to the
causal diamond), since this already suffices to illustrate the
problem.  In Sec.~\ref{sec-discussion} we will argue that this
limitation is an important criterion distinguishing the observations
made in a decelerating FRW universe from the S-matrix of
asymptotically flat space.

\section{Temperature and entropy of accelerating universes}
\label{sec-temp}

In this section we obtain the basic thermodynamic properties of
Q-space: entropy, energy, and temperature. We demonstrate that they
satisfy the first law of thermodynamics.  We begin by reviewing the
thermodynamics of de~Sitter space.

\subsection{Thermodynamics in de~Sitter space}
\label{sec-dstemp}

De Sitter space has an event horizon of radius $R_0=\sqrt{3/\Lambda}$.
Its area is $A=4\pi R_0^2=12\pi/\Lambda$.  It is also a Killing
horizon with surface gravity $\kappa=R_0^{-1}$, with respect to the
the usual timelike Killing vector field normalized at the origin.

Consider an object of mass $M$ in an otherwise empty asymptotically de
Sitter universe.  In the presence of this object, the cosmological
horizon will be smaller than that of empty de Sitter space.  One way
to estimate its size is to model the object as a small black hole.
For this case an exact solution is known: the Schwarzschild-de Sitter
black holes, with metric
\begin{equation}
ds^2= -V(r) dt^2+\frac{dr^2}{V(r)}+r^2 d\Omega_2^2\ ,
\end{equation}
where
\begin{equation}
V(r)=1-\frac{2M}{r}-\frac{\Lambda}{3} r^2\ .
\end{equation}
For $0<M<1/(3\sqrt{\Lambda})$, $V(r)$ has two positive roots.  (The
maximal case $M=1/(3\sqrt{\Lambda})$ is known as the Nariai solution;
larger black holes do not exist in de Sitter space.)  The smaller root
is the black hole horizon; it obeys $R_{\rm B}\approx 2M$ for small
$M$.  The larger root is the cosmological event horizon.  For small
$M$, it obeys
\begin{equation} 
R_{\rm C}^2\approx R_0^2-2R_0 M\ ,
\label{eq-rc}
\end{equation}
and it decreases monotonically over the whole range of $M$.

Now suppose that a black hole, or any other object of small mass $M$,
falls across the cosmological horizon, restoring the observer's patch
to empty de~Sitter space.  (This can be achieved simply by the
observer moving away from the object.)  By Eq.~(\ref{eq-rc}), this
process increases the cosmological horizon area by $\Delta A=-8\pi
R_0\, M$.  Thus, the cosmological horizon satisfies the usual first
law of horizon dynamics~\cite{BarCar73}:
\begin{equation}
-dE = \frac{\kappa\, dA}{8\pi}\ ,
\label{eq-bhd}
\end{equation}
where we have defined $dE$ to be the change in the mass of the matter
present on the observer's side of the horizon.\footnote{In black hole
  mechanics~\cite{BarCar73}, $dE$ thus corresponds to the change in
  mass of matter remaining {\em outside\/} the black hole, which is
  minus the change of black hole mass, and hence is negative when
  matter is added to the black hole.  Hence Eq.~(\ref{eq-bhd}) takes
  the same form for black holes and for de Sitter space.}  As in the
case black holes, this classical relation betrays the semiclassical
thermodynamic properties exhibited by the de~Sitter horizon.  Analysis
of quantum field theory in a de Sitter
background~\cite{GibHaw77a,BirDav} shows that a freely falling
detector will measure a temperature proportional to the surface
gravity
\begin{equation}
T=\frac{\kappa}{2\pi}~.
\end{equation}
Moreover, in order to avoid a decrease of observable entropy in the
above process, it is natural to propose that the horizon area
represents a true contribution to the total entropy, as originally
suggested for black holes~\cite{Bek72}:
\begin{equation}
S=\frac{A}{4}~.
\end{equation}
Consistency requires that these quantities satisfy the first law of
thermodynamics, which is ensured by Eq.~(\ref{eq-bhd}).

\subsection{Thermodynamics in Q-space}
\label{sec-qtemp}

Slow-roll inflation can be thought of as a de~Sitter-like era with
slowly decreasing effective cosmological constant.  It is well-known
that the apparent cosmological horizon during inflation has
thermodynamic properties akin to those of de Sitter
space~\cite{GibHaw77a}.  Indeed, this temperature is the origin of
density fluctuations and so is responsible for all structure in the
universe.

One would expect similar considerations to apply to an eternally
accelerating universe with $w$ sufficiently close to $-1$.  In this
case, the universe is also locally similar to de~Sitter space, with
slowly decreasing vacuum energy.  In fact, since a $w>-1$ fluid can be
modeled by a scalar field with custom-designed
potential~\cite{RatPee88}, it can be thought of as a special case of
slow-roll inflation.  Thus, horizon thermodynamics should apply in
Q-space.

We will now verify this expectation.  Our arguments will be rigorous
only for $w$ very close to a cosmological constant:
\begin{equation}
0<w+1 \ll 1~,
\end{equation}
though we expect our results to be qualitatively correct at least in
the range $-1<w<-2/3$.  The parameter $\epsilon = {3\over 2}(w+1)$
will be small and positive for the accelerating universes studied
here.  However, all classical formulas below are exact in $\epsilon$.

The radius of the event horizon was given in Eq.~(\ref{eq-re}).  We
will also be interested in the apparent horizon.\footnote{On each
  constant time slice, the apparent horizon of an observer at $r=0$ is
  the sphere whose orthogonal ingoing future-directed light-rays have
  vanishing expansion.}  In any FRW universe, its proper radius is
directly related to the energy density:
\begin{equation}
R_{\rm A} = \left(\frac{3}{8\pi\rho}\right)^{1/2}~.
\label{eq-rarho}
\end{equation}
For a flat universe, the apparent horizon radius is thus equal to the
Hubble scale, $t_{\rm H}=a/\dot{a}$, and is given by
\begin{equation}
R_{\rm A}=t_{\rm H} = \epsilon t\ .
\label{eq-rat}
\end{equation}

The two horizons satisfy the following key properties.  First, they
are approximately equal in the regime we study.  More precisely, the
apparent horizon is smaller than the event horizon by a fixed ratio
close to unity:
\begin{equation}
\frac{R_{\rm A}}{R_{\rm E}}= 1-\epsilon\ ,
\end{equation}
Second, neither horizon changes significantly over one Hubble time:
\begin{equation}
\frac{t_{\rm H} \dot{R}_{\rm X}}{R_{\rm X}}
 = \epsilon \ll 1;~~~\mbox{X=A,E}~.
\end{equation}
Hence, a thermodynamic description of the horizon will be
approximately valid, and it will not matter much whether we use the
apparent or the event horizon for this purpose.

We will work with the apparent horizon, since this approach is more
general.  (For example, in slow-roll inflation, there may be no event
horizon, but one would still like to describe the approximate thermal
state during inflation.)  Hence, an observer at $r=0$ will perceive a
thermal heat bath with slowly time-dependent temperature
\begin{equation}
T= \frac{1}{2\pi R_{\rm A}}
\label{eq-temp}
\end{equation}
and will ascribe to the apparent horizon a Bekenstein-Hawking entropy
\begin{equation}
S= \pi R_{\rm A}^2~.
\label{eq-entropy}
\end{equation}

As a consistency check, let us verify that the first law of
thermodynamics is satisfied.  We follow Ref.~\cite{BouHaw96}, where a
similar check was performed for slow-roll inflation.  Consider an
infinitesimal time interval $dt$.  The amount of energy crossing the
horizon during this time is obtained by integrating the flux of the
stress tensor across the surface, contracted with the (approximate)
generators of the horizon, the future directed ingoing null vector
field $k^a$:
\begin{equation}
-dE = 4\pi R_{\rm A}^2 \, T_{ab} k^a k^b \, dt
   = 4\pi R_{\rm A}^2 \, \rho (1+w) \, dt
   = \epsilon \, dt~.
\end{equation}
In the last equality we have used Eq.~(\ref{eq-rarho}).  By
Eqs.~(\ref{eq-entropy}) and (\ref{eq-rat}), the horizon entropy
increases by
\begin{equation}
dS = (2\pi R_{\rm A})\, \dot{R}_{\rm A} dt
   = (2\pi R_{\rm A})\, \epsilon\, dt~.
\end{equation}
The term in parentheses is the inverse temperature,
Eq.~(\ref{eq-temp}).  Thus we confirm the first law,
\begin{equation}
-dE=T\, dS~.
\end{equation}

\section{Thermal fluctuations in accelerating universes}
\label{sec-fluc}

Both in Q-space and in de Sitter space, the thermal horizon produces
fluctuations---but as we shall see in this section, their implications
are quite different in the two cases.  Fluctuations in de~Sitter space
are fatal to experiments.  We show, however, that in Q-space
fluctuations are benign: entropic enough to produce complex systems,
but not energetic enough to destroy an observer measuring them.

\subsection{Typical quanta}
\label{sec-typ}

We begin by asking: What is the total number of quanta emitted by the
horizon?  For de Sitter as for Q-space, the expected rate is one
quantum (typically with wavelength of order $R_{\rm A}$), per Hubble
time $R_{\rm A}$.  In de~Sitter space, $R_{\rm A}=R_0$ is a constant,
and an infinite number of quanta are emitted in total.  (No observer
will last long enough to notice more than a finite number, however, as
we shall see shortly.)  In Q-space, we see from Eq.~(\ref{eq-rat})
that $R_{\rm A}$ grows linearly with time.  However, the integrated
number of quanta still diverges (though only logarithmically, not
linearly as in the de Sitter case):
\begin{equation}
\int \frac{dt}{R_{\rm A}} \sim \log t \to \infty~.
\end{equation}

What is the total energy radiated?  In de Sitter space, the typical
energy of each quantum is fixed, so the radiated power integrates to
infinite energy, suggesting that it will erode any physical structure.
Any observer in de Sitter space will be thermalized by the steady
stream of radiation from the horizon.  

In Q-space, the rate of emission of quanta and the energy per quantum
each go like $R_{\rm A}^{-1}$.  Hence, the radiated power drops off
like the inverse square of time, and it integrates to a finite total
radiated energy.  Quantitatively the total energy radiated after the
time $t=t_0$ is
\begin{equation}
E \approx \int_{t_0}^\infty \frac{dt}{R_{\rm A}^2} 
  = \frac{1}{\epsilon} \int_{R_{\rm A}(t_0)}^\infty 
\frac{dR_{\rm A}}{R_{\rm A}^2}
  = \frac{1}{\epsilon R_{\rm A}(t_0)}~.
\end{equation}
For example, taking $t_0$ to be the time at which dark energy began to
dominate the evolution of our universe, the total energy (to be)
radiated by the cosmological horizon would be comparable to that of a
single quantum with wavelength of order the present Hubble scale.
Thus, the Q-space horizon falls far short of thermalizing the matter
it contains, in stark contrast with the de Sitter horizon.

\subsection{Large energy fluctuations in de~Sitter space}
\label{sec-dsenergy}

What is the probability for a state of specified energy $E$ to be
radiated by the horizon?  Aside from a slow death by thermalization,
observers in de Sitter space also face the threat of collisions with
objects of greater energy than the typical Hawking quanta.  Though
exponentially suppressed, such objects will eventually appear as rare
fluctuations in the thermal spectrum.  A particularly destructive
example is that of a nearly maximal Schwarzschild-de Sitter black
hole, which will swallow the observer.  

For small energy, $E\ll R_0$, the problem is approximately equivalent
to that of a hot cavity at temperature $T=(2\pi R_0)^{-1}$.  The
horizon provides the heat bath.  For larger energies, gravitational
backreaction can change the volume of the cavity and the temperature
of the horizon by factors of order unity.  In particular, there is a
largest possible energy, corresponding to a black hole that just fits
inside the cosmological horizon.  We will take these finiteness
effects into account but we begin by considering small energies.

The probability to find in the cavity a particular state $|i\rangle$,
of energy $E_i$, is given by
\begin{equation}
P(|i\rangle) = \frac{1}{Q}\, e^{-E_i/T}
 = \frac{1}{Q}\, \exp(-2\pi E_i R_0)~.
\label{eq-pi}
\end{equation}
For a cavity with radius of order the inverse temperature (and a
reasonable number of species), we can neglect factors of the partition
function,
\begin{equation}
Q=\sum_{|i\rangle}  \exp(-E_i/T)~,
\end{equation}
since it is dominated by a few states of energy $T$ and so is of order
unity.  Note that the probability $P(|i\rangle)$ is really a rate per
time interval of order the interaction time of the heat bath, $R_0$.

The probability to find an arbitrary state with energy $E$ is larger
than (\ref{eq-pi}) by a factor of the number of such states,
$N(E)=e^{S(E)}$:
\begin{equation}
\frac{P_E}{R_0} = \exp[S(E)-2\pi E R_0]~.
\label{eq-pe}
\end{equation}
A de Sitter space variant of the Bekenstein bound~\cite{Bek74,Bek81},
the D-bound~\cite{Bou00b}, guarantees that the exponent will be
non-positive.  For high energies compared to the thermal energy
$R_0^{-1}$, the second term in the exponent is large.  Thus the
rate of the corresponding fluctuations will be exponentially
suppressed, unless the entropy enhancement factor $e^{S(E)}$ nearly
cancels the suppression term, leaving an exponent of order unity.  We
now estimate $S(E)$ to argue that this is not the case.

In a quantum field theory coupled to gravity (a description which
should be locally valid at late times), the objects of highest entropy
for a given energy $E$ are either a black hole, or a radiation gas
with temperature $\tau$ and radius $\chi$ such that
$E\approx\chi^3\tau^4$.  (We assume that the number of species with
mass less than $\tau$ is not significant, i.e., less than $10^4$).
The entropy of the black hole is of order $E^2$.  The entropy of the
thermal radiation is $\chi^3\tau^3\approx (E\chi)^{3/4}$.  This is
maximized by choosing the radius occupied by the gas as large as
possible, $\chi=R_0$.  Thus the maximal entropy of thermal
radiation is
\begin{equation}
S_{\rm therm} \approx (ER_0)^{3/4}~.
\end{equation}
Whether this is larger than the black hole entropy $E^2$ depends on
the size of the horizon.

For $R_0^{-1}\lesssim E\lesssim R_0^{3/5}$, thermal
radiation wins.  Hence, in this regime we obtain the following upper
bound for the logarithm of the production rate:
\begin{equation}
S(E)-2\pi E R_0 \leq -2\pi (1-\delta) E R_0
\label{eq-exp}
\end{equation}
for some small number $\delta \approx (E R_0)^{-1/4}\ll 1$.  At
the level of accuracy required below, the $S(E)$ term can clearly be
dropped altogether ($\delta\approx 0$).

For $R_0^{3/5}\lesssim E\lesssim R_0$, a black hole dominates the
ensemble.  Near the lower end of this range, the black hole will have
radius $R_{\rm B}\approx 2E$.  For larger energy, however, the
backreaction on the cosmological horizon is significant, and the
definition of energy itself becomes ambiguous.  We will simply use the
black hole radius, $R_{\rm B}$, as an energy-like parameter and
abandon the estimate (\ref{eq-pe}) in favor of a direct computation of
the rate of black hole nucleation~\cite{Cha97,BouHaw98}:
\begin{equation}
\frac{P_{\rm B}}{R_0} = \exp[S_{\rm SdS}(R_{\rm B})-S_{\rm dS}]~.
\label{eq-pr}
\end{equation}
$S_{\rm SdS}(R_{\rm B})$ is the total entropy of a Schwarzschild-de
Sitter geometry with a black hole of radius $R_{\rm B}$.  It is given
by a quarter of the sum of the black hole and the cosmological horizon
area.  $S_{\rm dS}=\pi R_0^2$ is the entropy of the empty de Sitter
solution with the same cosmological constant.  Einstein's equation
implies for any static spherically symmetric vacuum
solution~\cite{BouHaw98}:
\begin{equation}
R_{\rm B}^2+R_{\rm C}^2+R_{\rm B} R_{\rm C}= R_0^2~.
\label{eq-radii}
\end{equation}
Here, $R_{\rm C}$ is the radius of the cosmological horizon.  Hence
the creation rate (\ref{eq-pr}) is simply
\begin{equation}
\frac{P_{\rm B}}{R_0} = \exp[-\pi R_{\rm B} R_{\rm C}]~.
\label{eq-pr2}
\end{equation}
The exponent agrees well with Eq.~(\ref{eq-exp}) in a large region of
overlap: For $R_{\rm B}\ll R_0$, one can take $R_{\rm B}\approx 2E$.
Moreover, the contribution $S(E)$ from the black hole entropy is
subleading.

Already the smallest black holes, with $R_{\rm B}\approx 1$ and
$P_{\rm B}\sim\exp(-\pi R_0)$, are exponentially suppressed and thus
very unlikely to arise in the thermal spectrum.  At fixed cosmological
constant, one finds from Eq.~(\ref{eq-radii}) that $R_{\rm C}(R_{\rm
  B})$ is a monotonically decreasing function: the cosmological
horizon gets smaller for larger black holes.  But $R_{\rm B} R_{\rm
  C}(R_{\rm B})$ grows monotonically, so larger black holes are more
and more unlikely.  The biggest black hole allowed by
Eq.~(\ref{eq-radii}) has $R_{\rm B}=R_0/\sqrt{3}$ and is suppressed by
$\exp(-\pi R_0^2/3)$.

However, no matter how small the rate of such fluctuations, in de
Sitter space it is independent of time.  Hence, even the most unlikely
fluctuation will eventually occur, on a timescale of order $R_0/P$.

\subsection{Large energy fluctuations in Q-space}
\label{sec-qenergy}

In a $w>-1$ accelerating universe, Eqs.~(\ref{eq-pi}), (\ref{eq-pe}),
and (\ref{eq-pr}) still describe the probability for the corresponding
fluctuations, if we substitute $R_{\rm A}$ for $R_0$.  But as we shall
see now, violent events of a specified energy $E$ are not likely to
ever occur at late times, no matter how long one waits.

First consider a fluctuation of less than the Planck energy, $E\leq
1$.  Its rate is given by Eq.~(\ref{eq-pe}).  Let $t_0$ be a
sufficiently late time so that the temperature of the horizon has
become small compared to the energy of the fluctuation: $R_{\rm
  A}(t_0)=\epsilon t_0\gg E^{-1}$.  What is the total probability
${\cal P}(E)$ for the fluctuation of energy $E$ to occur after the
time $t_0$?
\begin{eqnarray} 
{\cal P}(E) & = & \int_{t_0}^{\infty} 
dt\,\frac{P_E(t)}{R_{\rm A}(t)}\\
& \leq & \frac{1}{R_0}\int_{t_0}^\infty 
dt\, \exp[S(E)-2\pi ER_{\rm A}(t)].\\
& \leq & \frac{1}{\epsilon R_0}\int_{R_{\rm A}(t_0)}^\infty 
dR_{\rm A}\, 
\exp[-(2\pi-\delta) ER_{\rm A}]\\
& = & \frac{1}{(2\pi-\delta)\epsilon ER_{\rm A}(t_0)}
\exp[-(2\pi-\delta) ER_{\rm A}(t_0)].
\end{eqnarray}
(Here $\delta \approx [E R_{\rm A}(t_0)]^{-1/4}$.)  Since $ER_{\rm
  A}(t_0)\gg 1$, the total probability is exponentially small.

Fluctuations greater than the Planck energy cannot be considered until
the horizon has grown large enough to contain a black hole of energy
$E$.  During the period $E/\epsilon\lesssim t\lesssim
E^{5/3}/\epsilon$, a fluctuation of energy $E$ is most likely to occur
in the form of a black hole.  But this power-law time interval is
insufficient to overcome the exponential suppression in
Eq.~(\ref{eq-pr}), so the fluctuation is extremely unlikely to occur
during this period.  Thereafter, the thermal ensemble begins to
dominate, and the fluctuation rate is given by Eq.~(\ref{eq-pe}).
Then the analysis of the previous paragraph applies, with
$t_0=E^{5/3}/\epsilon$.  Since $ER_{\rm A}(t_0)$ is again large, the
integrated probability remains negligible for all times.

We conclude that large energy fluctuations inevitably occur in de
Sitter space (if only after an exponentially large time), guaranteeing
the destruction of any observer.  In Q-space, however, the temperature
falls monotonically.  After it drops below a given energy $E$,
fluctuations of that energy become virtually impossible.

\subsection{Large entropy fluctuations in de~Sitter space}
\label{sec-dsentropy}

What is the probability for a state of specified entropy $S$ to be
radiated by the de Sitter horizon?  We have seen in the previous
subsection that the probability of fluctuations is mainly determined
by their energy; the entropy factor in Eq.~(\ref{eq-pe}) turned out to
be negligible.  Hence the question is, what is the lightest object
with entropy $S$?

For $S\lesssim R_0^{6/5}$ the lightest object is a thermal state with
temperature $\tau\approx S^{1/3}/R_0$ and energy $E\approx
S^{4/3}/R_0$.\footnote{At least it is the lightest object that we are
  sure exists.  If a lighter object has the same entropy,
  Eqs.~(\ref{eq-pst}), (\ref{eq-pps}) and (\ref{eq-ppps}) still
  provide lower bounds on its rate of production, leaving our
  conclusions intact.}  It is radiated with a probability derived from
Eq.~(\ref{eq-pe}):
\begin{equation}
\frac{P_S}{R_0} = \exp(S-\alpha S^{4/3})~.
\label{eq-pst}
\end{equation}
(Here $\alpha$ is a numerical coefficient involving Stefan's constant
and the effective number of species with mass below $\tau$; for small
species number, it will be on the order of $10$.)  For $S\gtrsim
R_0^{6/5}$ the lightest object is a black hole with radius
$R_{\rm B}=(S/\pi)^{1/2}$.  Thus Eq.~(\ref{eq-pr2}) implies
\begin{equation}
\frac{P_S}{R_0} = \exp[-(\pi S)^{1/2} R_{\rm C}]~.
\label{eq-psb}
\end{equation}
From Eq.~(\ref{eq-radii}) it follows that the suppression becomes
stronger if $S$ is increased at fixed cosmological constant.  Since
these rates are constant, the situation is similar to the case of
large-energy fluctuations: All events that can occur in de~Sitter
space, will occur.

However, there is an absolute entropy bound in de~Sitter
space~\cite{Ban00,Bou00a}.  There are no states with entropy greater
than that of the horizon of empty de~Sitter, $\pi R_0^2$.  This bound
refers to the combined entropy of the cosmological horizon and of the
matter it encloses.  If we ask about the entropy only of systems
contained {\em within\/} the horizon, the limit is more stringent:
there can be no objects with entropy greater than that of the Nariai
black hole ($\pi R_0^2/3$).  Fluctuations with greater entropy cannot
occur; their probability is exactly zero.  This fundamentally limits
the complexity and accuracy of experiments in de Sitter space.

\subsection{Large entropy fluctuations in Q-space}
\label{sec-qentropy}

In Q-space the horizon grows linearly.  A fluctuation of entropy $S$
first becomes possible (in the form of a maximal black hole) when the
horizon reaches a radius of order $S^{1/2}$.  Thus, the rate begins at
$e^{-S}$, the suppression of a Nariai black hole.  Thereafter the
required black hole radius remains constant.  But the corresponding
radius of the cosmological horizon, $R_{\rm C}$, increases as the
effective cosmological constant decreases, according to
Eq.~(\ref{eq-psb}).  Hence the fluctuation becomes more and more
unlikely.  Eventually the horizon radius satisfies
\begin{equation}
R_{\rm A}\gtrsim S^{5/6}~.
\end{equation}
In this regime, the lightest object of entropy $S$ is an ordinary
thermal state.  For all remaining time, the rate of such a fluctuation
is given by Eq.~(\ref{eq-pst}).  

What matters about this asymptotic rate is not its (miniscule) value,
but that it is both constant and non-zero, however large one chooses
$S$.  It depends on $R_{\rm A}$ only in that $R_{\rm A}$ is the time
interval for which $P_S$ represents the probability of one
fluctuation.  Hence the integrated probability ${\cal P}$ for a
fluctuation of entropy $S$ diverges logarithmically:
\begin{eqnarray} 
{\cal P}(S) & = & \int_{t_0}^{\infty} dt\,\frac{P_S}{R_{\rm A}}
\label{eq-pps}\\
& \leq & \epsilon^{-1} \exp(S-\alpha S^{4/3})
\int_{S^{5/6}}^\infty \frac{dR_{\rm A}}{R_{\rm A}} \to\infty.
\end{eqnarray}

This is a remarkable result: objects of any complexity, no matter how
large, will eventually be emitted by the horizon.  The key observation
is that at fixed entropy, there are many ``scaling states'' whose
energy is inversely proportional to their linear size.  As the horizon
grows, these states become energetically cheaper at the same rate as
the temperature drops, leaving their probability invariant.  This
restricts consideration to massless fields at late times.\footnote{In
  our estimates, this restriction is implemented by using the number
  of {\em massless\/} species in the thermodynamic formulas for the
  energy and entropy of a thermal cavity.  Note that other types of
  universes will also have only massless particles at late times, if
  massive particles are unstable or processed by black holes.}

Indeed, the stronger statement holds that each scaling microstate
individually is produced with certainty:
\begin{equation}
{\cal P}(|i\rangle) = \epsilon^{-1}\exp(-\alpha S^{4/3})
\int \frac{dR_{\rm A}}{R_{\rm A}} \to\infty~.
\label{eq-ppps}
\end{equation}
This includes highly structured, irregular configurations.

\section{Asymptotic observables}
\label{sec-discussion}

In this section we will compare the various cosmological solutions
studied above, with an eye on the complexity and precision of
measurements that can be achieved, and on the possibility of defining
exact asymptotic observables or an S-matrix.  

By an asymptotic observable, we mean any quantity that can be measured
with arbitrary precision at sufficiently late times.  We expect that
asymptotic observables exist only in spacetimes where experiments of
arbitrarily long duration can be made and an arbitrarily large amount
of entropy can be accessed.  An S-matrix is a special case of an
asymptotic observable, consisting of matrix elements between the
complete initial and final asymptotic states of a closed isolated
system.

The limitations on observation discussed here are imposed by
fundamental aspects of the cosmological solution, such as its causal
and thermal properties, and its information content.  This allows us
to proceed without any assumptions about the nature of experiments or
observers.\footnote{Additional restrictions may arise, for example,
  from a limited supply of free energy or inability to harvest this
  energy for experiments (see, e.g.,
  Refs.~\cite{Dys79,Wit01,KraSta00,BusAda03}).  To the extent that
  they are insurmountable, they may further constrain the asymptotic
  observables.}

\subsection{De~Sitter}

Asymptotically de Sitter spacetimes ($w=-1$) are particularly hostile
to observers.  There is no S-matrix, since the observer's causal
diamond misses almost all of the asymptotic regions, $\eta\to 0$ and
$\eta\to\pi$ in the global metric (\ref{eq-global}), on which the
global in and out-states might be defined.  An S-matrix between such
states would, at best, be a ``meta-observable''~\cite{Wit01}: It would
relate a state no one can set up (because of the past event horizon
$\eta=\chi$) to a state no one can measure (because of the future
event horizon $\eta=\pi-\chi$).  This conclusion does not improve if
the past asymptotic region is replaced by a big bang; this only trades
the past event horizon for a particle horizon, and does not affect the
future event horizon.

Nor are there any other asymptotic observables in asymptotically
de~Sitter space.  The total accessible entropy is bounded
(Sec.~\ref{sec-class}), and the duration of any experiment
fundamentally limited by thermal erosion (Sec.~\ref{sec-typ}) and
by collisions with black holes (Sec.~\ref{sec-dsenergy}).

\subsection{Q-space}

Like de~Sitter space, quintessence dominated universes ($-1<w<-1/3$)
have a cosmological event horizon~\cite{HelKal01,FisKas01} (see
Sec.~\ref{sec-frw}).  Hence, the global state in the asymptotic future
cannot be measured, and there is no S-matrix.

However, some other asymptotic observables may well exist.  We have
seen in the previous section that Q-space is significantly more
welcoming to physicists than de~Sitter space.  Thermal
fluctuations\footnote{We should emphasize again that the thermal
  properties of Q-space discussed in Secs.~\ref{sec-temp} and
  \ref{sec-fluc} were rigorously derived only in the limit of small
  $\epsilon$ ($w\to -1$).  But we expect no qualitative transitions at
  least in the range $-1<w<-2/3$.}  are present but are too weak to
terminate experiments by erosion (Sec.~\ref{sec-typ}) or by black hole
production (Sec.~\ref{sec-qenergy}).  Because the cosmological horizon
becomes arbitrarily large~\cite{HelKal01}, there is no absolute
entropy bound.  What we showed in Sec.~\ref{sec-qentropy} is that an
unbounded number of different states are actually produced at late
times.  Thus, observers can experience arbitrarily complex events
(and, one might imagine, store large amounts of information for long
times).\footnote{Whether these fluctuations, which involve only
  massless fields, can give rise to an apparatus capable of precise
  measurements is another question, and we do not claim to have proven
  that this will happen.}

\subsection{Decelerating FRW}

Decelerating universes ($-1/3<w<1$) clearly satisfy important
conditions for the existence of asymptotic observables, as noted by
several authors~\cite{BanFis01a,HelKal01,FisKas01,Wit01,FreSus04}.  As
the particle horizon grows, the amount of entropy allowed in the
causal diamond increases without bound, as does its actual matter
content (Sec.~\ref{sec-class}).  This may include massive particles,
if they are stable and if they are not converted to radiation by black
holes.

But is there an S-matrix?  At first sight, the
situation looks promising.  There is no future event horizon
(Sec.~\ref{sec-frw}).  Every timelike geodesic eventually enters the
causal diamond of the observer.  But this does not mean that the
global state of the universe is observable.

At any finite time, only a finite portion of the universe is in the
observer's causal diamond, by Eq.~(\ref{eq-finite}).  Beyond lies a
non-compact region, which has {\em at all times\/} infinite volume (as
measured on the homogeneous spacelike slices).  This unobserved region
contains an infinite amount of matter and, potentially, an infinite
amount of information.  Thus, the decelerating universe never reveals
more than an infinitely small fraction of itself to the observer (see
Fig.~\ref{fig-penfrw}).

Whether all or only part of a system is measured, makes an enormous
difference.  Page~\cite{Pag93} has shown that in order to obtain at
least one bit of information about a system in a typical pure state,
one must perform a measurement on more than half of the degrees of
freedom constituting the system.  Thus, even a measurement of one half
of a finite system by the other half will reveal practically no
information whatsoever about the global state.  The situation in a
flat FRW universe is far more problematic yet: the number of degrees
of freedom available for measurement are finite, and the total system
is infinite.

Let us compare this to a real S-matrix experiment, such as the
scattering of particles in an accelerator.  Here, the entire system
hits the detector by some finite time.\footnote{The fact that the
  complete system is observable does not rely on any limiting
  procedure.  The usual limit of late times and large detectors is
  taken only in order to refine the separation of particles and make
  sure that they have stopped interacting.  This is a separate
  requirement related to our preference for expressing the out-state
  in a convenient Hilbert space basis (the Fock space of the
  noninteracting theory).}  The key difference is that in
asymptotically flat space, there exists a region near spatial infinity
(i.e., outside a sufficiently large sphere) that is devoid of matter
and energy.  Entropy bounds, such as the Bekenstein bound and the
generalized covariant bound~\cite{Bek81,FMW,Bou03}, imply that this
region contains no information.  In an FRW solution, on the other
hand, the density at fixed time is asymptotically constant and
non-zero.  In this case, entropy bounds permit an arbitrarily large
information content.  Therefore, the asymptotic structure of an FRW
universe does not guarantee that an S-matrix exists.

The situation could be improved by restricting to a set of states such
that the (infinite) exterior of some finite region contains either no
information or only redundant information.  (See, e.g.,
Ref.~\cite{FreSus04} for an approach to constructing an appropriate
reference state.)  We emphasize that this is a strong additional
constraint.  The appropriate states would form a set of measure zero
in the Hilbert space of states of the FRW universe.  It is possible,
but not obvious, that suitable states are selected as initial
conditions by theory.

\subsection{Discussion: Cosmology vs.\ the S-matrix}

We have found that the entropy of observable matter is unbounded in
any flat FRW universe dominated by a $w>-1$ fluid in the asymptotic
future, accelerating or not.  In particular, we conclude that a future
event horizon does not in itself impose a significant restriction.
Its absence is neither necessary for the existence of asymptotic
observables nor sufficient for the existence of an S-matrix.  Indeed,
we argued that an S-matrix is not a natural observable in any of the
cosmologies considered.

We based our argument on the combination of two observations: the
late-time global state of the universe is never fully contained in any
observer's causal past; and unlike asymptotically flat space, the
unobserved portion of a cosmological universe can contain
information---in the case of decelerating universes, it contains an
infinite number of degrees of freedom.  Page's theorem~\cite{Pag93}
then implies that no information can be gleaned about a generic global
pure quantum state, if by information we mean finding a density matrix
of sub-maximal entropy.

This is just a particularly bad version of a more general problem that
arises whenever one part of a closed system measures another part.
This includes any measurement of the global state of the universe,
independently of causal restrictions.  Obviously, the apparatus must
have at least as many degrees of freedom as the system whose quantum
state it attempts to establish (in practice it usually has orders of
magnitude more).  This means that at most half of the degrees of
freedom can be observed, just below the Page cutoff for obtaining the
first bit of information about the complete system.

Aside from the problem of measuring the out-state, an S-matrix
description of cosmology is emptied of operational meaning by our
inability to control the initial state and to repeat experiments that
extend over cosmological time and distance scales.

It is conceivable that these problems could be
circumvented.\footnote{I would like to thank L.~Susskind for
  discussions of this issue.}  Suppose for example that theory
restricts to states with a high degree of spatial symmetry and no
entanglement between distant degrees of freedom.  In an infinite
universe without event horizons, it might then be possible to perform
independent but equivalent measurements on an arbitrary number of
disentangled identical subsystems.  At all times, however, the
construction of a global state would still require infinite
extrapolation.  In any case, this procedure will not resemble an
S-matrix experiment.

If not an S-matrix, what asymptotic observables should one expect to
find?  In any large spacetime, local high-energy scattering
experiments will admit an S-matrix description to a good
approximation.  But we are interested in measurements probing the
state and the evolution of the universe.  We are fortunate to inhabit
a fairly symmetric (part of the) universe, and we can learn aspects of
its early quantum fluctuations by measuring correlations in the cosmic
microwave background.  They are limited by cosmic variance, but as our
horizon grows we can obtain more and more data points.  In
decelerating universes such correlations may become exact quantum
observables in the asymptotic future.  In accelerating universes, one
would measure correlations in fluctuations of the approximately
thermal radiation from the horizon.

Whether or not asymptotic observables exist, they do not correspond
precisely to the experiments we perform {\em today}.  Even in an
asymptotically empty spacetime (suppose, for instance, that our
universe is a mere resonance in a giant scattering event in
asymptotically flat space), one would require approximate local
observables to describe measurements by parts of the system on one
another at finite time.  They would seem just as likely to be
computable from an initial state directly, than to be derived from
S-matrix.  (Indeed, the former option has a chance of applying in
general spacetimes, whereas the latter requires assumptions about the
asymptotic structure.)  We do not even know whether such bulk
observables are quantum mechanical.

\section{Other universes}
\label{sec-other}

The discussion above was restricted to flat FRW universes, which may
admit some asymptotic observables though not, in any operational
sense, an S-matrix.  In this section, we extend the discussion to
other universes.  We focus in particular on the Farhi-Guth solution,
which connects a cosmological region, through the interior of a black
hole, to an asymptotically flat or AdS spacetime, raising anew the
question of S-matrix observables for cosmology.

\subsection{Crunching universes}

Many other cosmological solutions do not admit
asymptotic observables at all.
This includes all universes with a big crunch, such as closed FRW
solutions with decelerating matter content.  Open and flat FRW
solutions can also crunch, even if they are initially expanding, if
they contain negative vacuum energy~\cite{KalLin99}.  In this case,
spatial infinity exists.  However, the largest causally accessible
region has finite entropy~\cite{CEB1,CEB2}, so there will be no exact
observables.

There are also spacetimes that do not crunch globally but in which all
observers must end up inside a black hole.  In a dust-dominated flat
universe, for example, gravitational collapse can occur at arbitrarily
large scales.  Our own universe was probably produced by a period of
inflation.  Then the assumption of an infinite homogeneous flat FRW
universe is not appropriate at very large scales.  Assuming a generic
chaotic inflation potential, the density fluctuations produced by
inflation grow logarithmically with the scale.  Exponentially many
years from now, the fluctuations entering the horizon will be of order
unity.  On some scales this will lead to large voids, on other scales
to overdensities.  Sooner or later, any given observer will find
themselves in a large region bound to collapse into a black hole.

\subsection{Asymptotically expanding, non-flat FRW universes}

Open universes have a similar causal structure to flat universes.  The
main difference is that in the accelerating (decelerating) cases, the
future conformal boundary is isomorphic only to a portion of the
infinity of de Sitter (Minkowski) space.  Hence, the spatial infinity
of an open universe is a conformal two-sphere rather than a point.

The discussion of observables in flat FRW universes, in
Sec.~\ref{sec-discussion}, applies to open universes as well.  In the
accelerating case ($w<-1/3$), this is because the spatial curvature
vanishes at late times anyway.  In the decelerating case, asymptotic
observables may exist, since the particle horizon diverges; but the
global state of the universe is again unobservable.  Closed Q-space
also asymptotes locally to flat Q-space, as the curvature inflates
away.

\subsection{Coleman-De Luccia}

\begin{figure}
  \includegraphics[width=8.5cm]{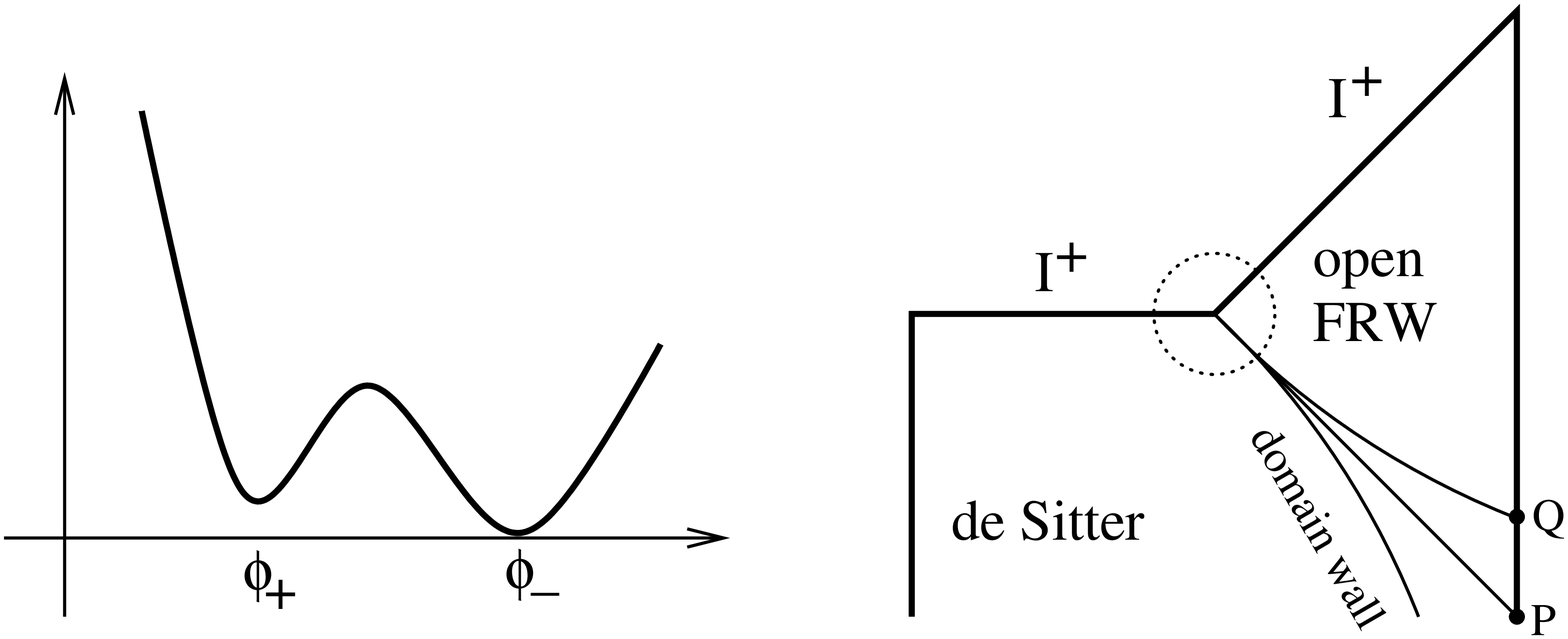}
\caption{\label{fig-penfrwd}
  A scalar field potential involving a false vacuum (left) gives rise
  to the Coleman-De Luccia solution, which describes a de Sitter
  region joined to an open FRW universe, in our example with vanishing
  vacuum energy.  A conformal diagram of the expanding portion of this
  spacetime is shown on the right.  Some examples of the orbits of the
  symmetry group $O(3,1)$ are shown: the domain wall, the light-cone
  starting at $P$, and the hyperbolic time slice that includes
  $Q$.---The region enclosed in the dotted circle is asymptotically
  identical with a corresponding region of the Farhi-Guth solution
  (Fig.~5).}
\end{figure}

One might hope to gain control over a poorly behaved universe, such as
de Sitter space, by grafting onto it another solution which possesses
asymptotic observables or perhaps even an S-matrix.  An example of
such a hybrid is the Coleman-De Luccia (CDL) solution~\cite{ColDel80},
which describes a bubble of true vacuum expanding inside a false
vacuum.  Of particular interest is the case where the true vacuum
energy vanishes (so the false one is positive).  The expanding portion
of this solution describes an open expanding FRW universe joined to a
meta-stable de Sitter space (Fig.~\ref{fig-penfrwd}).  Unlike pure
de~Sitter space, which appears to allow no exact observables, the CDL
solution should therefore admit the same asymptotic observables that
can be defined in an open FRW universe.\footnote{One can also consider
  the fully extended CDL solution, which contains a collapsing FRW
  universe in the past.  Freivogel and Susskind~\cite{FreSus04}
  recently proposed that asymptotic observables defined in the two FRW
  regions encode information not only about the metastable de Sitter
  region, but even about other stable and metastable
  vacua~\cite{BouPol00,KKLT} in causally disconnected regions.  This
  raises a number of subtle questions~\cite{BouFre05} which are
  outside the scope of this paper.}

It is important that the $\Lambda=0$ region of the CDL solution is an
FRW cosmology~\cite{ColDel80} and not, as one might have expected,
empty Minkowski space.  That is, it contains infinite hyperbolic
slices with constant positive energy density.  For later use, we
briefly review where this energy comes from.

The true and false vacuum can be modeled by a scalar field potential
with a local minimum at $\phi_+$ and a global minimum at $\phi_-$
(Fig.~\ref{fig-penfrwd}).  The domain wall of the CDL solution is a
spherical shell in which the field $\phi$ crosses the barrier between
the vacua.  Consider the closed, time-symmetric slice on which the
domain wall radius takes its minimum value.  If the energy of the
false vacuum is small compared to the height of the barrier, then the
thickness of the domain wall will be small compared to its radius.
This limit is known as the thin-wall approximation.

Inside the wall, the field value is approximately $\phi_-$, differing
from the exact vacuum value only by an amount exponentially small in
the distance from the wall.  The wall is only of finite size, so at
its center (at the event $P$ in Fig.~\ref{fig-penfrwd}) the field
value $\phi_0$ still differs from $\phi_-$ by an exponentially small
amount.  (This can be avoided only by infinite fine-tuning of the
potential.)  Hence, the energy density at $P$ is not exactly zero.  By
continuity, the energy density will also be nonzero at some point $Q$
infinitesimally later than $P$.  The $O(3,1)$ symmetry of the CDL
solution guarantees that $Q$ is equivalent to all other events on the
spatial hyperbolic slice generated as its orbit.  Hence, the entire
infinite slice has constant positive energy density.  This is what
distinguishes a cosmology from asymptotically flat space.

By the same token, if the true vacuum has negative cosmological
constant, the bubble interior will not be Anti-de Sitter space.  It
will again be an open FRW solution~\cite{ColDel80,Ban02}. Like any FRW
solution with negative cosmological constant and an admixture of
$w>-1/3$ matter~\cite{KalLin99}, it will only expand for a finite
amount of time and then collapse in a big crunch.

\subsection{Farhi-Guth}

The Coleman-DeLuccia solution is a limiting case of a larger class of
domain wall solutions with spherical symmetry, found by Blau,
Guendelman, and Guth~\cite{BlaGue87} (see also
Refs.~\cite{AurDen85,BerKuz83}).  This class includes another
composite cosmology, the Farhi-Guth solution\footnote{We refer to it
  by the authors of Ref.~\cite{FarGut87}, who investigated features of
  this solution, to distinguish it from the larger class of which it
  is a special case.}, which does contain a true asymptotically flat
region.  This suggests that it may allow the description of cosmology
using an S-matrix.  A similar solution can be constructed with
Anti-de~Sitter asymptotics; it has been suggested that aspects of the
cosmological regions could thus be described via the AdS/CFT
correspondence~\cite{AlbLow99}.\footnote{I have enjoyed discussions
  with B.~Freivogel, V.~Hubeny, M.~Rangamani, and S.~Shenker, who are
  independently investigating questions that overlap with some of the
  topics in this subsection~\cite{FreHub05}.}

This possibility has met with some scepticism (see, e.g.,
Ref.~\cite{Ban02}).  Here we demonstrate a feature of the Farhi-Guth
solution which has not, to our knowledge, been previously noted in the
literature: the fact that it contains an open, asymptotically FRW
universe with a black hole.  We will argue that this exacerbates the
difficulties with using the Farhi-Guth solution for an S-matrix
description of cosmology.

\paragraph*{Global structure} 

\begin{figure}
  \includegraphics[width=8.5cm]{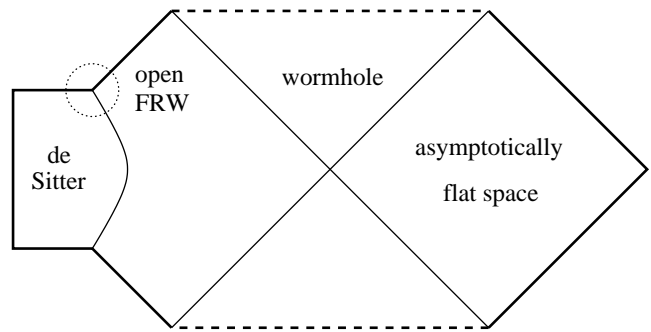}
\caption{\label{fig-penfrwe}
  Conformal diagram of the fully extended Farhi-Guth solution.  In the
  text it is shown that the region marked by the dotted circle agrees
  asymptotically with the marked region in Fig.~4; this implies that
  it contains an open universe.  The above diagram corresponds to the
  case of vanishing cosmological constant in the true minimum; for
  negative cosmological constant, the same argument shows that the
  open universe crunches.}
\end{figure}

The Farhi-Guth solution describes an expanding bubble of de Sitter
space topologically ``inside'' an asymptotically flat universe.  The
only way~\cite{FarGut87} this can be achieved is to place the de
Sitter bubble on the far side of a black hole/white hole region, as
shown in Fig.~\ref{fig-penfrwe}.  To allow for quick orientation, let
us call this the ``cosmological side'', separated by an Einstein-Rosen
bridge from the ``asymptotic side''.  Points on opposite sides are
necessarily spacelike separated, so one cannot travel between them.
Note that the cosmological side is very similar to the CDL solution:
it contains a meta-stable de~Sitter region separated by a domain wall
from a region of vanishing cosmological constant.

At late times, the Farhi-Guth bubble grows large, and its wall will be
far from the black hole.  Thus, it can be expected to behave
asymptotically like the CDL domain wall.  We will now verify this.
The dynamics of the Farhi-Guth wall is governed by the Israel junction
conditions, which yield the equation~\cite{FarGut90}
\begin{equation}
(\frac{dr}{d\tau}) ^2 + V(r,q) =-1\ .
\end{equation}
Here, $r$ is the radius of the bubble, $\tau$ is the proper time on
the bubble trajectory, and $q$ is the radius of the black hole.  The
potential $V(r,q)$ is given by
\begin{equation}
V(r,q) = -\frac{q}{r} - 
\frac{[q-(\chi^2+\kappa^2) r^3]^2}{4\kappa^2 r^4}~,
\end{equation}
where $\chi$ is the Hubble scale of the meta-stable de~Sitter region,
and $\kappa/4\pi$ is the surface tension of the domain wall.  These
latter quantities are determined by the shape of the scalar field
potential.

The CDL solution corresponds to setting $q=0$, so $V(r)= -
\frac{(\chi^2+\kappa^2)^2 r^2}{4\kappa^2}$.  This admits a growing and
a decaying exponential solution.  The particular, time-symmetric
initial conditions of CDL select the linear combination
\begin{equation}
r = r_0 \cosh \frac{\tau}{r_0}~,
\end{equation}
where $r_0 = 2\kappa/(\chi^2+\kappa^2)$~.

The Farhi-Guth solution is more complicated but we are interested only
in the large radius limit.  For $r\to\infty$ one finds that $V(r,q)\to
- \frac{(\chi^2+\kappa^2)^2 r^2}{4\kappa^2}$, which coincides with the
$q\to 0$ limit.  Hence, growing bubbles are all attracted to the CDL
solution at late times, independently of the black hole mass:
\begin{equation}
r \to C \exp \frac{\tau}{r_0}~,
\end{equation}
Differences in the prefactor can be absorbed into a shift of the time
variable $\tau$.  

This universality has an important consequence: the formation of an
expanding FRW universe outside the bubble will also be universal.  In
the CDL case, we exploited the full $O(3,1)$ symmetry to show that the
domain wall dumps constant energy density into the hyperbolic slices.
The details of this mechanism, and the character of the matter it
produces, will vary depending on the potential and couplings.  The
Farhi-Guth solution has only spherical symmetry.  But at large radius
the initial conditions for the hyperbolic slices are provided mainly
by the domain wall.  They are identical to those in the CDL solution,
so the same mechanism will operate, creating asymptotically hyperbolic
slices with asymptotically constant, positive energy density.

Hence, the region of vanishing vacuum energy on the cosmological side
of the Farhi-Guth solution is not empty flat space but is again a
cosmology.  It is asymptotic to an open FRW universe.  The only
difference to the CDL solution is that the asymptotic FRW universe
contains a black hole, though which it connects to an asymptotically
flat region.

\paragraph*{Discussion}

The Farhi-Guth solution has it all: an asymptotically flat or AdS
region, meta-stable de~Sitter space, a black hole, an open FRW
cosmology, and in the case of AdS asymptotics, even a big crunch.
Could it be that all these interesting regions are described by an
S-matrix, or CFT correlators, defined on the asymptotic
side?\footnote{See Refs.~\cite{Ban02,Ban04} for other discussions of
  this issue.}

The region behind the horizon is causally inaccessible to an observer
at infinity.  To argue that information about the cosmological side
can be retrieved on the asymptotic side, one would have to appeal
either to black hole complementarity~\cite{AlbLow99} or to subtle
effects of analyticity~\cite{KraOog03,FidHub04}.  But complementarity
is a stronger conjecture here than in the case of classical black hole
formation, since the matter on the cosmological side was never present
on the asymptotic side.  

Let us restate this in the language of holographic
screens~\cite{CEB2}.  For a black hole formed in asymptotically flat
space by the collapse of matter, the past null infinity is a screen
whose light-sheets reach inside the black hole, covering all of the
spacetime.  They are appropriate for a description of the infalling
observer; all information that went into the black hole can
unambiguously be stored there.  If this trivial fact was not true,
then the assumption that the S-matrix is unitary would not require
that the same information be present at future infinity, and we would
not be led to complementarity.  In the Farhi-Guth solution, the
light-sheets off of the asymptotic boundary enter the black hole/white
hole region, but they cannot reach into the cosmological regions on
the far side of the black hole.

A signal that complementarity may not work here is the fact that the
black hole area can be made arbitrarily small, while the de~Sitter
region on the cosmological side can have large entropy.  This
contrasts with the usual case, where the black hole area grows large
in response to matter crossing the horizon, and becomes small only as
it returns matter to the asymptotic region in the form of Hawking
radiation.

The infinite open FRW universe, which we have argued is always present
on the cosmological side, exacerbates the entropy mismatch, leaving
little hope that it could be resolved by some unknown macroscopic
constraint on the solutions.  The cosmological side would have to be
in one of a small number of very special microstates, if it were to
be described by boundary data defined on the asymptotic side.  Another
logical possibility would be that the boundary theory makes more
degrees of freedom available for the cosmological side of the black
hole than it does for the black hole itself; this would lead to a
breakdown of the UV/IR correspondence of AdS/CFT~\cite{SusWit98}.

These entropic considerations are complemented by an aesthetic
objection: If nature had wanted us to use a boundary theory to
describe cosmology, it would have given the universe a nicer boundary.
After all, the Farhi-Guth solutions are rather artificial constructs.
Their description reads like a cocktail recipe: A de Sitter region
separated by a domain wall from an open universe containing an eternal
black hole, on the far side of whose Einstein-Rosen bridge resides the
desired asymptotically flat region.  (The idea of the Einstein-Rosen
bridge is logically independent of the de~Sitter region; for example,
one could connect any FRW universe to an asymptotic region this way.)

Unlike the CDL solution, which arises naturally from the decay of
false vacuum, spacetimes containing an eternal black hole are not of
obvious physical relevance.  They cannot be created classically from
regular initial conditions~\cite{FarGut87}.  It has been argued that
the Farhi-Guth geometry might arise
semiclassically~\cite{FarGut90,FisMor90a,FisMor90b} from a small
bubble of false vacuum, but no regular instanton exists for this
process.  Moreover, it is not clear how an observer would distinguish
this type of transition from other exponentially rare events resulting
in the spontaneous formation of a black hole.

In summary, it is questionable whether Farhi-Guth solutions exist in a
full quantum theory of gravity; and even if they do, holographic
considerations suggest that the asymptotic and the cosmological side
will be completely decoupled.  That said, we are unable to rule out
that some aspects of cosmological evolution are encoded in boundary
data via the Farhi-Guth solution.

\paragraph*{Acknowledgements} 

I would like to thank T.~Banks, B.~Freivogel, V.~Hubeny, N.~Kaloper,
A.~Linde, A.~Mints, J.~Polchinski, M.~Rangamani, S.~Shenker,
L.~Susskind, and E.~Witten for discussions.  This work was supported
by the Berkeley Center for Theoretical Physics, by a CAREER grant of
the National Science Foundation, and by DOE grant DE-AC03-76SF00098.

\bibliographystyle{board}
\bibliography{all}
\end{document}